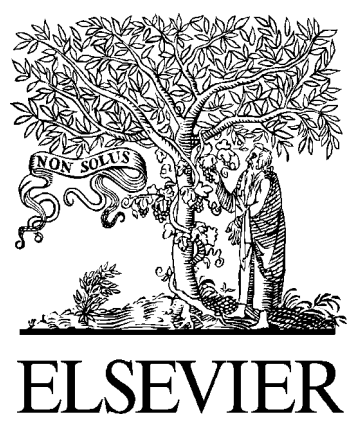

# Optimal Microgrid Sizing of Offshore Renewable Energy Sources for Offshore Platforms and Coastal Communities

Ann Mary Toms[a], Xingpeng Li[a], Kaushik Rajashekara[a]

[a]University of Houston, Department of Electrical and Computer Engineering, Houston, TX, 77204-4005

**Abstract**

The global energy landscape is undergoing a transformative shift towards renewable energy and advanced storage solutions, driven by the urgent need for sustainable and resilient power systems. Isolated offshore communities, such as islands and offshore platforms, which traditionally rely on mainland grids or diesel generators, stand to gain significantly from renewable energy integration. Promising offshore renewable technologies include wind turbines, wave and tidal energy converters, and floating photovoltaic systems, paired with a storage solution like battery energy storage systems. This paper introduces a renewable energy microgrid optimizer (REMO), a tool designed to identify the optimal sizes of renewable generation and storage resources for offshore microgrids. A key challenge in such models is accurately accounting for battery degradation costs. To address this, the REMO model integrates a deep neural network-based battery degradation (DNN-BD) module, which factors in variables like ambient temperature, charge/discharge rates, state of charge, depth of discharge and battery health. Simulations on six test regions demonstrate that the REMO-DNN-BD approach minimizes lifetime energy costs while maintaining high reliability and sustainability, making it a viable design solution for offshore microgrid systems.

Nomenclature

Abbreviations

| | | | |
|---|---|---|---|
| *DNN-BD* | Deep neural network-based battery degradation | *FPV* | Floating photovoltaic systems |
| *OREM* | Offshore renewable energy microgrid | *OWT* | Offshore wind turbines |
| *REMO* | Renewable energy microgrid optimizer | *TEC* | Tidal energy converters |
| *WEC* | Wave energy converters | | |

Sets

| | | | |
|---|---|---|---|
| *F* | Set of all capacities of marine grade FPV systems in kW | *O* | Set of all capacities of OWT systems in kW |
| *H* | Set of Hours in a day | *T* | Set of all capacities of TEC systems in kW |
| *M* | Set of Months in a year | *W* | Set of all capacities of WEC systems in kW |

Indices

|  |  |  |  |
|---|---|---|---|
| $f$ | Electric Power Capacity of the marine-grade FPV | $o$ | Electric Power Capacity of OWT |
| $h$ | Hour of the day | $t$ | Electrical Power Capacity of TEC |
| $i$ | Iteration Counter | $w$ | Electrical Power Capacity of WEC |
| $m$ | Month of the year | | |

Parameters

|  |  |  |  |
|---|---|---|---|
| $C_{BESS}^{capital}$ | Capital cost of each kWh of BESS. | $C_f^{precom}$ | Pre-commissioning cost of a single marine grade FPV unit of $f$ kW Capacity. |
| $C_f^{capital}$ | Capital cost of a single marine grade FPV unit with a capacity of $f$ kW. | $C_o^{precom}$ | Pre-commissioning cost of a single OWT unit of $o$ kW capacity. |
| $C_o^{capital}$ | Capital cost of a single OWT unit with a rated capacity of $o$ kW. | $C_t^{precom}$ | Pre-commissioning cost of a single TEC unit of $t$ kW Capacity. |
| $C_t^{capital}$ | Capital cost of a single TEC unit with a capacity of $t$ kW. | $C_w^{precom}$ | Pre-commissioning cost of a single WEC unit of $w$ kW capacity. |
| $C_w^{capital}$ | Capital cost of a single WEC unit with a rated capacity of $w$ kW. | $P_{Max}^{Char}$ | Maximum Charging Power of BESS. |
| $C_{BESS}^{decom}$ | Decommissioning cost for each kWh of BESS. | $P_{Max}^{Disc}$ | Maximum Discharging Power of BESS. |
| $C_f^{decom}$ | Decommissioning cost of a single marine grade FPV unit with a capacity of $f$ kW. | $P_{m,h}^{f}$ | Available solar power generated from the $f$ kW marine grade FPV at month $m$ at hour $h$ |
| $C_o^{decom}$ | Decommissioning cost of a single OWT unit of $o$ kW capacity. | $P_{m,h}^{Load}$ | Load at month $m$ at hour $h$ . |
| $C_t^{decom}$ | Decommissioning cost of a single TEC unit with a capacity of $t$ kW. | $P_{m,h}^{o}$ | Available wind power generated from the $w$ kW OWT at month $m$ at hour $h$ . |
| $C_w^{decom}$ | Decommissioning cost of a single WEC unit of $w$ kW capacity. | $P_{m,h}^{t}$ | Available tidal power generated from the $t$ kW TEC at month $m$ at hour $h$ . |
| $C_{BESS}^{O\&M}$ | O&M cost of each kWh of the BESS unit per annum. | $P_{m,h}^{w}$ | Available wave power generated from the w kW WEC at month $m$ at hour $h$ . |
| $C_f^{O\&M}$ | O&M cost associated with each marine grade FPV unit of $f$ kW per annum. | $SOC_{min}$ | Minimum state of charge of BESS. |
| $C_o^{O\&M}$ | O&M cost associated with a single OWT unit of $o$ kW per annum. | $SOC_{max}$ | Maximum state of charge of BESS. |
| $C_t^{O\&M}$ | O&M cost associated with each TEC unit of $t$ kW per annum. | $T_e$ | Expected lifetime of the OREM system. |
| $C_w^{O\&M}$ | O&M cost associated with a single WEC unit of $w$ kW per annum. | $\eta_{BESS}^{Char}$ | Charging Efficiency of the BESS. |
| $C_{BESS}^{precom}$ | Pre-commissioning cost of each kWh of BESS. | $\eta_{BESS}^{Disc}$ | Discharging Efficiency of the BESS. |

Variables

|  |  |  |  |
|---|---|---|---|
| $E_{BESS}$ | Energy capacity of a single BESS unit. | $N_o$ | Number of OWT units with a capacity of $o$ kW. |
| $E_{BESS}^{initial}$ | Initial energy capacity of the BESS unit. | $N_w$ | Number of OWT units with a capacity of $w$ kW. |
| $E_{BESS}^{total}$ | Total energy capacity of the BESS unit. | $P_{m,h}^{Char}$ | Charging power of BESS at month $m$ at hour $h$ |
| $E_{BESS}^{m,h}$ | Energy capacity of the BESS unit at month $m$ at hour $h$ . | $P_{m,h}^{Curt}$ | Renewable power curtailment at month $m$ at hour $h$. |
| $N_{BESS}$ | Number of 3900kWh BESS units | $P_{m,h}^{Disc}$ | Discharging power of BESS at month $m$ at hour $h$. |
| $N_t$ | Number of marine-grade TEC units with a capacity of $t$ kW. | $U_{m,h}^{Char}$ | Charging status of BESS at month $m$ at hour $h$. 1 represents charging mode and 0 represents discharging or idle mode. |
| $N_f$ | Number of marine-grade FPV units with a capacity of $f$ kW. | $U_{m,h}^{Disc}$ | Discharging status of BESS at month $m$ at hour $h$. 1 represents charging mode and 0 represents discharging or idle mode. |

## 1. Introduction

Coastal, remote and islanded communities face unique energy challenges due to their geographic isolation, which affects energy accessibility, affordability and reliability. Climate change exacerbates these issues through rising sea levels, environmental threats and more frequent and extreme storms. Most of these communities rely on electricity from the mainland and fossil fuel based energy sources [1]. However, as climate change intensifies, continued reliance on these traditional resources will inevitably result in rising energy costs. According to the U.S. Energy Information Administration (EIA), as of 2023, the island state of Hawaii has one of the highest electricity rates in the country, averaging 38.60 cents/kWh [2], significantly above the national average of 12.68 cents/kWh [3]. Similarly, many remote rural communities in Alaska that lack interconnection to the main transmission grid face even higher electricity costs, ranging from 50 to 80 cents/kWh [4].

### *1.1. Literature Review*

Offshore renewable energy resources like wave energy converters (WEC), tidal energy converters (TEC), offshore wind turbines (OWT), and floating photovoltaic systems (FPV), offer a promising alternative to the conventional energy sources for these regions. These technologies have significant untapped potential and are experiencing rapid growth due to technological advancements [5]. The U.S. Department of Energy (DOE) estimates that the theoretical annual energy potential from wave resources along the U.S. coasts ranges between 1,594 and 2,640 TWh, while tidal resources could provide approximately 445 TWh annually [6]. Offshore wind energy also holds immense potential in the United States, with an estimated capacity exceeding 2 TW [7]. OWTs are notably larger than their onshore counterparts, benefiting from stronger and more consistent winds at sea, which results in higher capacity factors for offshore wind farms [8]. Additionally, advancements in floating wind farms are enabling the development of projects in deep and ultra-deep waters, expanding the reach of offshore wind energy. Marine grade FPVs are a nascent technology and are still being researched. However, federally controlled reservoirs have a potential of 2.1 TW [9] using FPVs.

However, WECs face reliability and economic challenges due to changing ocean patterns[10], but array layouts [11] aligned with wave direction can enhance performance [12-13]. TECs offer stable output in high-density regions[14]. OWTs [15] and FPVs [16-17] are intermittent, increasing energy costs and complicating integration. Hybrid systems combining multiple renewables generation resources [18-20] and storage like batteries, ultracapacitors, fuel cells, and pumped hydro [21-24] can mitigate intermittency. Advances in AI-driven forecasting improve reliability [25-30].

Research on the optimal sizing of renewable microgrids explore a variety of methodologies, each with its own strengths and limitations. Commercial software like HOMER [31-32], are widely used because of their accessibility and interactive interfaces. However, they often rely on simplified assumptions regarding system dynamics which leads to oversizing of the microgrid [33]. Heuristic approaches like genetic algorithm [34-35], non-dominated sorting genetic algorithm [36-37], particle swarm optimization [24,38], are capable of handling multi-objective functions, but, they can suffer from convergence issues [39]. Newer bio-inspired methods, like the artificial jelly search [40], attempt to overcome these limitations through more dynamic exploration strategies, but their effectiveness in high-dimensional problems remains under investigation. Mathematical techniques like mixed integer linear programming [41,42], provide high precision and rigor for microgrid optimization problems. However, this approach often requires linear approximations and can become computationally intensive when modeling nonlinearities like battery degradation and renewable intermittency. Hybrid techniques [43] aim to combine the accuracy of mathematical programming with the adaptability of heuristics, often improving both convergence and realism. These studies also explore the optimization of different objective functions including reduced cost [24, 38, 42-52], reduced CO2 emission [35], reduced curtailment [37,38], improved reliability [24, 36-38] and stability [32, 35]. Additionally, to account for the intermittent nature of the renewable resources, various stochastic networks are employed including Monte Carlo simulations [44]. In contrast, our proposed model builds on these foundations by addressing key limitations that are often overlooked in traditional approaches. This allows for more realistic and resilient microgrid designs, particularly in offshore or remote settings where reliability and efficiency are critical.

BESS are widely used in islanded offshore microgrids with high renewable penetration [18,43-44], but hybrid energy storage systems have recently shown greater reliability and stability[47-48]. Offshore conditions demand BESS with high volumetric and gravimetric energy density, safety, long lifespan, and low maintenance [49-50]. Environmental factors like ambient temperature, humidity, mechanical stress, dust and debris accumulation [41, 51] use based factors like depth of charge/discharge, rate of charge, state of charge, and state of health of the BESS [52] significantly affect performance and lifespan. Studies have explored the electrochemical and mechanical properties of lithium-ion batteries in salt spray environments [53] and have found that salt spray exposure led to battery swelling due to imperfect sealing. Data-driven models [52], including Gaussian process regression and Kalman filters [54], have improved degradation prediction accuracy. These insights underscore the importance of accurate BESS degradation modeling for optimal energy storage system design and resource planning.

BESS degradation modeling has advanced to account for complex interactions between environmental and usage-based factors. Semi-empirical [55], physics-aware [56], mathematical [57], thermodynamic [58], and temporal [59] models have been developed to estimate BESS degradation and integrate into optimization frameworks. Several key factors influence BESS degradation including ambient temperature [57-60], cycling patterns [58-61], state of charge [62], thermal stress [60], chemical reactions [60] and natural aging [63]. Understanding these factors is vital for enhancing BESS reliability, performance, and lifespan. Table 1 shows a comparison of existing battery degradation models reported in literature.

Table 1. Comparison of existing battery degradation models reported in literature

| Ref. | Model Type | Input Parameters | Application | Strengths | Limitations |
|---|---|---|---|---|---|
| [52] | Fully Connected Neural Network | • Ambient Temperature<br>• Charge/discharge rate<br>• State of charge<br>• Depth of discharge<br>• State of health | Microgrid day ahead scheduling | Able to capture non-linear interaction between input parameters. | Battery aging test data is limited and can be leveraged only for Lithium ion batteries |
| [54] | Gaussian Process Regression and extended Kalman Filter combined algorithm | Voltage of the battery at different time points | Electrified vehicles | Reduced computational burden | Limited experimental data. Low degree of model nonlinearity |
| [55] | Semi-Emperical Model | • Battery calendar aging<br>• State of charge<br>• Ambient Temperature<br>• Battery cycle life<br>• Depth of discharge | Frequency regulation dispatch | Model combined theoretical analysis with experimental observation. Hence model is applicable over a range of operating conditions | The algorithm can only be applied to offline evaluations, is unsuitable for real time evaluations |
| [56] | Physics aware model | SEI growth parameters | Frequency regulation dispatch | Highly accurate estimates | High computation time |
| [57] | Mathematical Model | • Cycle count<br>• Temperature<br>• Charge/discharge rate<br>• State of charge<br>• Depth of discharge<br>• Capacity fade | Battery Life optimization | Accounts for non-linear degradation trajectory | Lack of application specific testing |
| [58] | Thermodynamic Model | • Terminal voltage<br>• Voltage drop across load resistor<br>• Battery temperature<br>• Ambient temperature | Battery modelling | Generalizable model for Lithium ion batteries of all chemistries | Lack of application specific testing |
| [59] | Temporal Model | • Charge Power<br>• Discharge Power<br>• Ambient Temperature | Predicting capacity fade | Real world data collected from behind the meter setup.<br>Provided real-time | Unified ageing framework may oversimplify the interaction between input parameters |

| | | • State of Charge<br>• State of Health | | suggestions to improve BESS lifetime | |
|---|---|---|---|---|---|

Existing battery degradation models present two key limitations that will be addressed in this paper. First, most models consider only a limited subset of degradation factors particularly, ambient temperature and state of charge, while overlooking other critical factors like state of health, cycling patterns, depth of discharge, natural ageing etc. This narrow focus significantly limits the accuracy and reliability of BESS degradation prediction, particularly in real world applications where multiple factors interact nonlinearly over time. Second, these models often assume simplified BESS operation conditions, failing to account for the highly dynamic nature of actual BESS operations. Neural networks have shown strong potential in improving degradation predictions. Accurate modeling is essential for optimizing performance and supporting effective planning and operational decisions in energy storage systems.

### *1.2. Objectives and Contributions*

The primary aim of this paper is to explore renewable marine energy generation resources like WECs, TECs, OWTs and FPVs as viable alternatives to fossil fuel based solutions. This paper also aims to advance the optimization of microgrid systems designed to integrate these renewable resources. Current microgrid design approaches often overlook the intermittent and seasonal nature of renewable resources [34,44] and oversimplify the system economics by only considering the capital and operational cost [38, 43-44, 47-48, 50]. Additionally, many sizing models optimize for a single fixed size of a generation resource, simplifying computations but risking under-sizing or over-sizing. Another common simplification is estimating BESS degradation costs as a fixed percentage of the capital cost, which may lead to inaccurate results, especially in systems with frequent charging and discharging cycles, where degradation is faster than anticipated.

This paper presents a Renewable Energy Microgrid Optimizer (REMO), a model that determines the optimal mix of renewable generation resources integrated into an offshore renewable energy microgrid (OREM) for independently powering coastal or island communities. The OREM integrates WECs, TECs, OWTs, FPVs, and BESS to meet to meet energy demands solely through renewable sources. The proposed REMO model optimizes the sizing of generation and storage resources, accounting for the stochastic nature of the renewable energy generation resources. Unlike conventional models, REMO evaluates multiple sizes of renewable generation resources and incorporates a deep neural network-based battery degradation (DNN-BD) model to precisely calculate costs associated with use-based battery degradation. The proposed DNN-BD model aims to improve the accuracy of degradation cost estimation in the OREM. The DNN-BD model captures the complex, nonlinear interaction amongst five key variables namely: ambient temperature, charge/discharge rate, state of charge, depth of discharge and state of health and is integrated into the REMO sizing framework, enabling more accurate and cost effective BESS planning.

The rest of this paper is organized as follows. Section 2 details the projection of electrical power outputs from the different generation resources. Section 3 details subsystem level modelling of the OREM. Section 4 discusses the integration of the neural network-based battery degradation model. In Section 5, an update to the sizing model is introduced to help reduce overall computation time. The test cases and associated simulations and results are presented in Section 6. The paper is concluded in section 7 with the future works discussed in section 8.

## 2. Projection of Electrical Power Generation from Different Offshore Renewable Resources

Renewable energy generation is inherently intermittent, making it crucial to accurately calculate the power output from different renewable sources. In this research, we used five years of historical data from datasets maintained by the National Data Buoy Center (NDBC) and the National Oceanic and Atmospheric Administration (NOAA) [64-65]. These datasets are used to estimate the average electrical power output from OWTs, TECs and WECs over a 24-hour period for each hour of each month for the following regions in the USA: Texas, California, Hawaii, Alaska, Florida, and New Jersey. Figure 1 pinpoints the locations and station IDs of the regions around which the different datasets were collected.

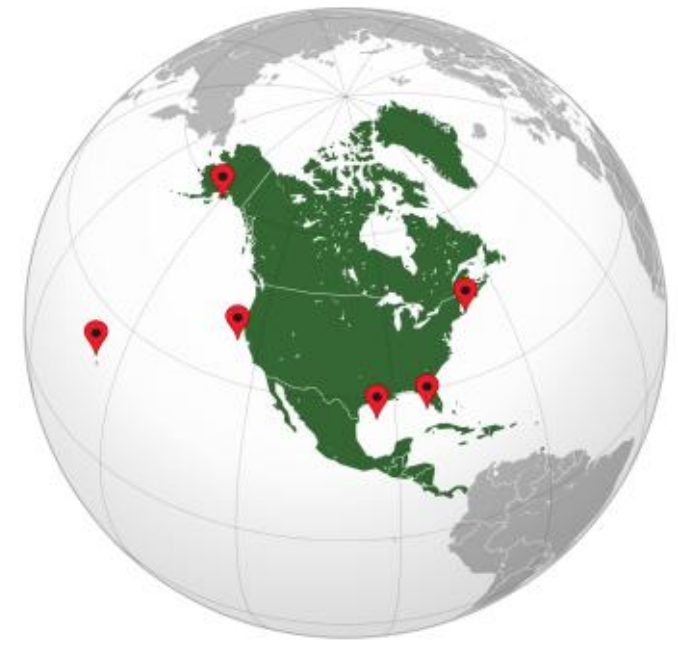

| *Region* | *NOAA Station ID* | *NDBC Station ID* |
|---|---|---|
| ***TX*** | G06010 | 42011 |
| ***AK*** | COI0519 | 46001 |
| ***CL*** | SFB1223 | 46011 |
| ***FL*** | FPI0901 | 41010 |
| ***HI*** | HAI1127 | 51000 |
| ***NJ*** | DEB2101 | 44091 |

Fig. 1. Geographical map along with station IDs of the six regions analyzed in this study: Texas, California, Hawaii, Alaska, Florida, and New Jersey.

## *2.1. Electrical Power Generation from WECs*

There are various types of WECs, but for this research, we focus on a linear attenuating WEC. This type of WEC harnesses wave energy by using the rise and fall of swells to create a flexing motion, which drives hydraulic pumps to generate electricity. The power matrix [66-67] is used to estimate the electrical power output of a single WEC. Figure 2 illustrates the average power output that a single WEC can generate for each hour of every month.

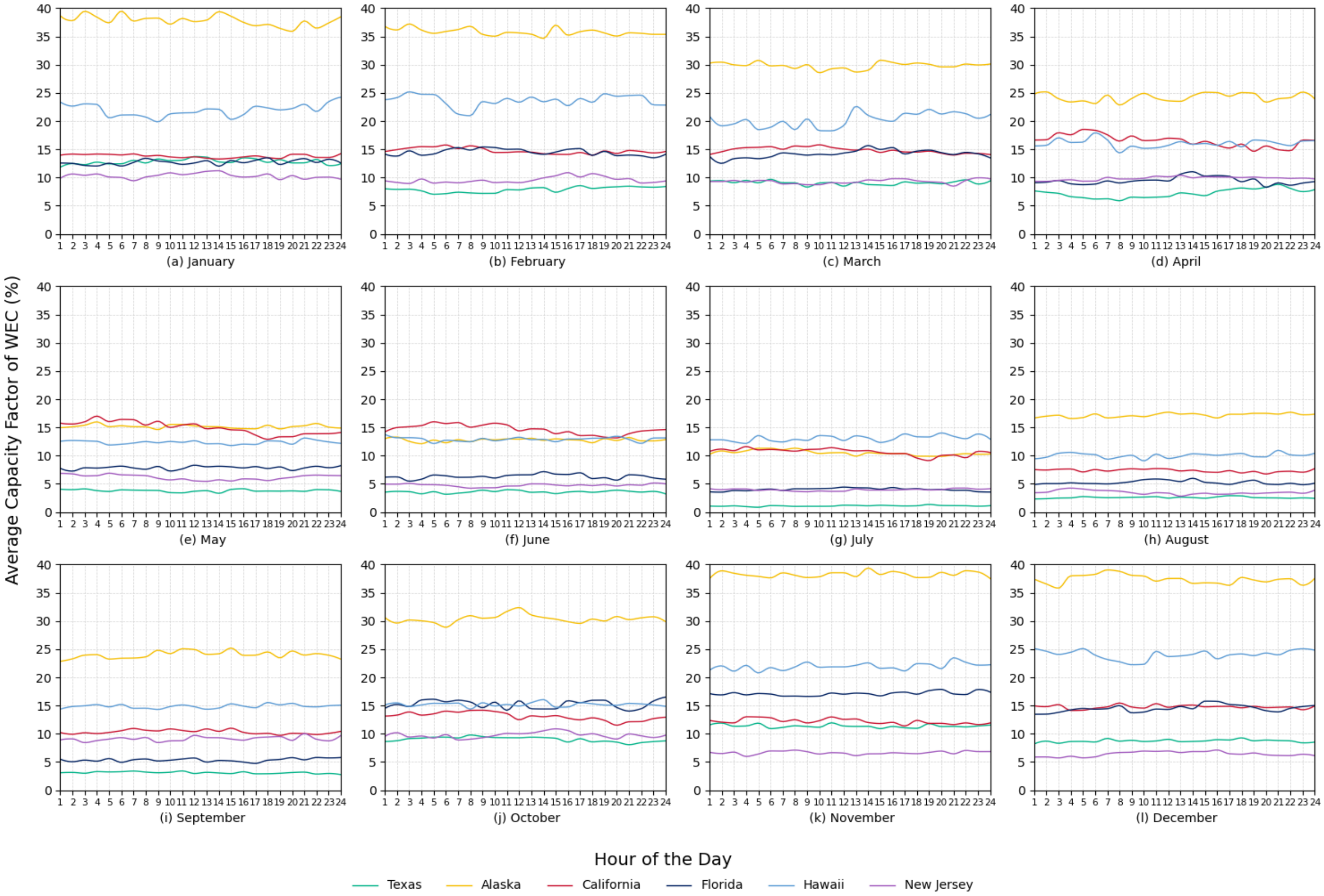

Fig. 2. Average output capacity factor of a single WEC for each hour of each month.

From the figure, it’s clear that highest power output occurs mainly during the winter months, from November to February for all of the six regions. WECs perform best in Alaska followed by Hawaii. The highest average capacity factor is observed in Alaska during January (subplot 2.a) reaching an average of 38%. However, as seen in subplot 2.g, this capacity factor drops significantly to just 10% outside of winter. In contrast, WEC performance in Texas is much lower, with a capacity factor of only 13% in January (subplot 2.a) and reaching its lowest point of just 1% in July (subplot 2.g).

### *2.2. Electrical Power Generation from TECs*

TECs generate electricity from water currents in the same way that OWTs generate power from air currents or wind. However, the power generation potential of a TEC can often surpass that of a similarly rated OWT. This is because water is denser than air, allowing a single TEC to produce significant power even at lower tidal flow velocities compared to wind speeds. Despite this potential, TECs are still in the early stages of development in the renewable energy market. The total electrical power that can be generated from a TEC is calculated using (1):

$$P_{m,h}^{t} = \frac{1}{2} \times \rho_t \times A_t \times \left(v_{m,h}^{t}\right)^3 \times C_{p_t} \times \eta_t \tag{1}$$

where $\rho_t$ is density of water where the TEC is installed; $A_t$ swept area of the TEC rotor; $v_{m,h}^{t}$ is the tidal velocity for month m at hour h; $C_{p_t}$ coefficient of power; and $\eta_t$ is efficiency of electrical system. Figure 3 illustrates the average power output that a single TEC can generate for each hour of every month.

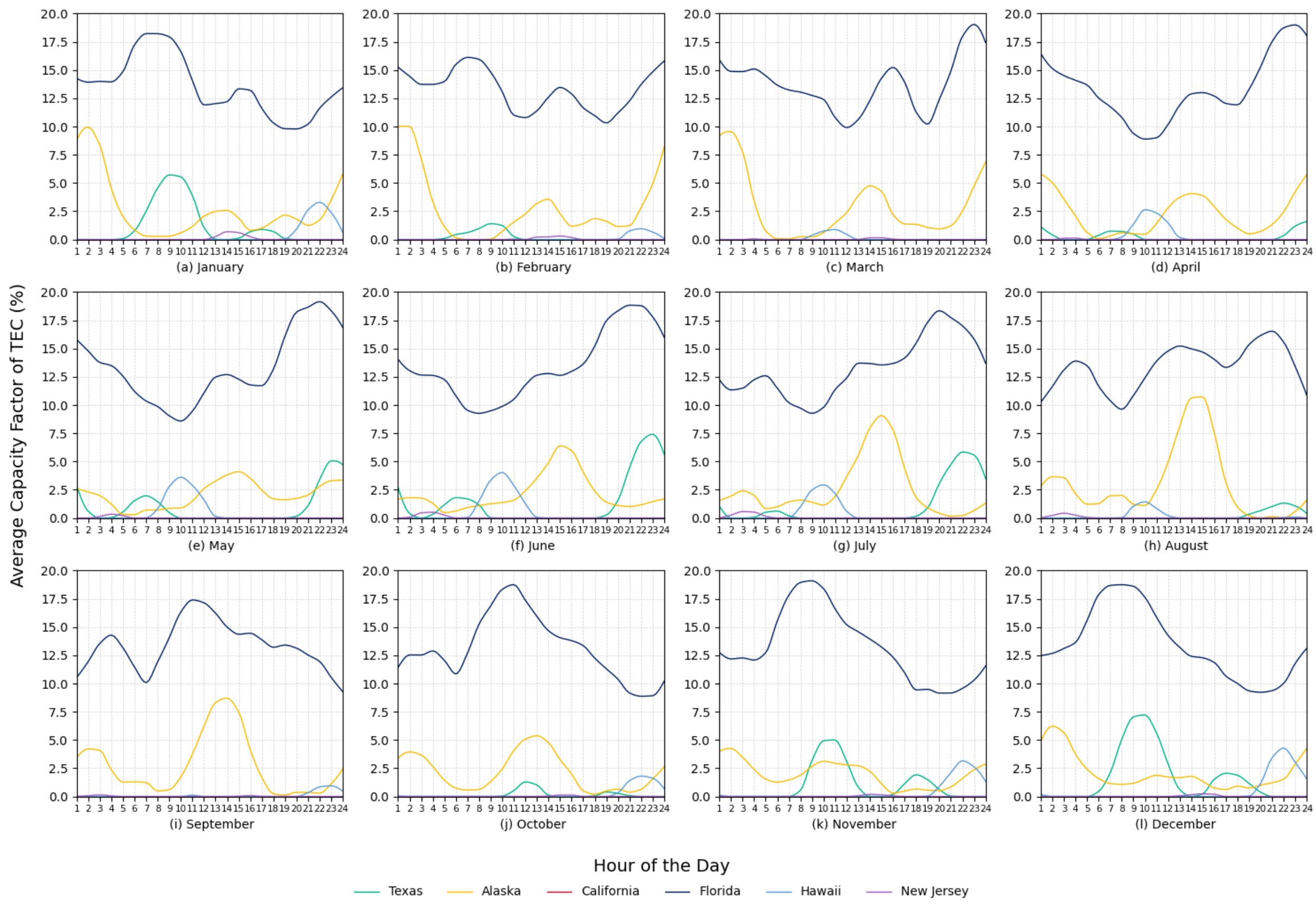

Fig. 3. Average output capacity factor of a single TEC for each hour of each month.

From the figure, it's clear TECs perform best in Florida followed by Alaska and almost no power can be generated from TECs in New Jersey and California. The highest average capacity factor is observed in Florida during May (subplot 3.e) reaching an average of 19%. In contrast, to WECs, the performance of TECs is constant throughout the year with an average capacity factor of 13% for all months in Florida.

### 2.3. Electrical Power Generation from OWTs

OWT technology is one of the most advanced renewable energy solutions available. The OWTs generate electricity by harnessing strong winds that rotate their blades, which then convert mechanical energy into electrical energy through the principle of electromagnetism. The total electrical power that can be generated from an OWT is calculated using (2):

$$P^{o}_{m,h} = \frac{1}{2} \times \rho_o \times A_o \times \left(v^{h_2}_{m,h}\right)^3 \times C_{p_o} \times \eta_o \qquad (2)$$

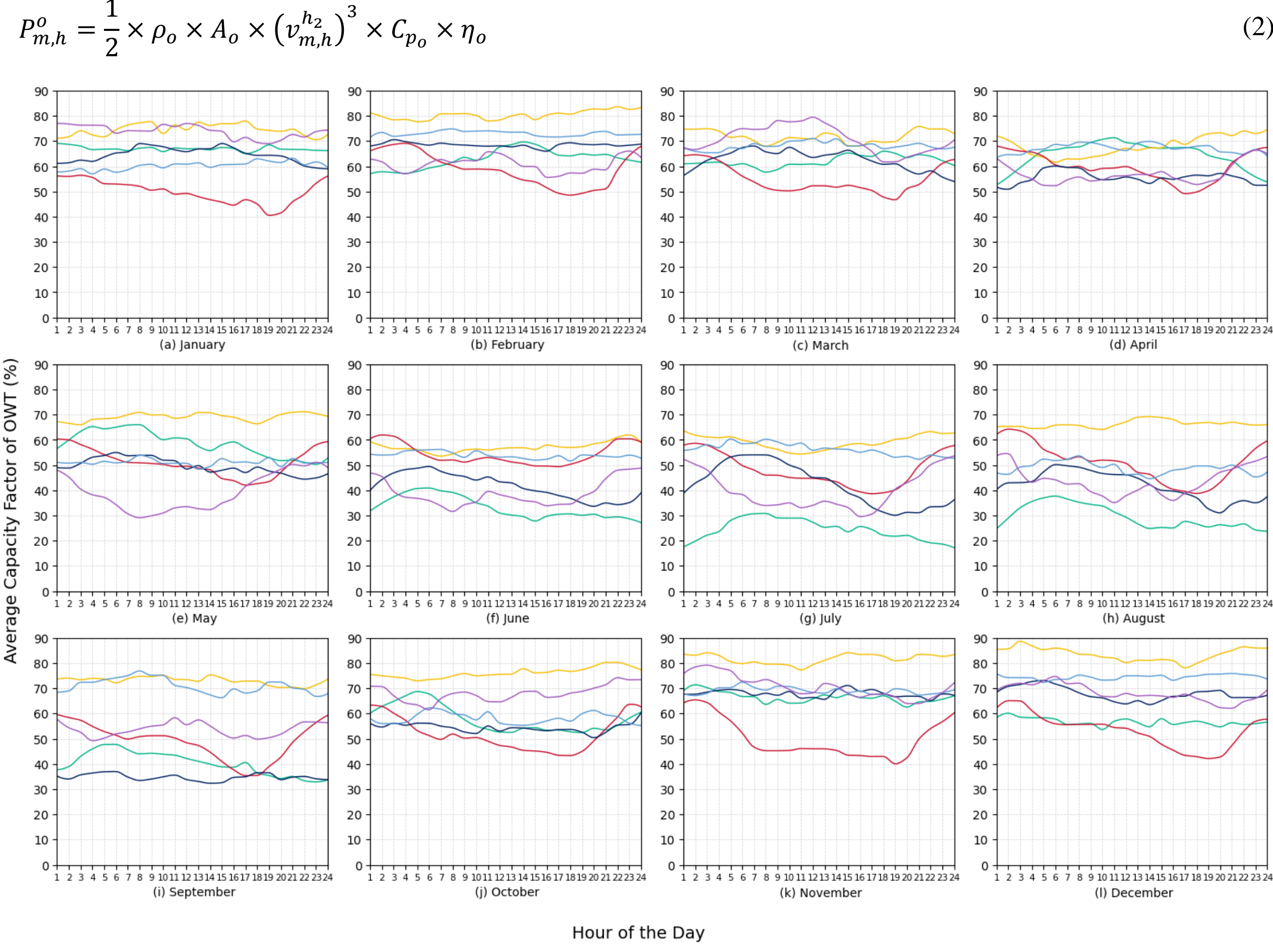

Fig. 4. Average output capacity factor of a single 8000kW OWT for each hour of each month.

where $\rho_o$ is density of air at the hub height; $A_o$ swept area of the OWT rotor; $v^{h_2}_{m,h}$ is velocity of wind at the hub height for month m at hour h; $C_{p_o}$ coefficient of power; and $\eta_0$ is efficiency of electrical system. Since most datasets do not provide wind speed measurements at desired hub height of around 80m for OWT, the log wind profile [30,68-69] is used to estimate wind velocity at this height, given by (3):

$$v^{h_2}_{m,h} = v^{h_1}_{m,h} \frac{ln\left(\frac{h_2}{z_0}\right)}{ln\left(\frac{h_1}{z_0}\right)} \qquad (3)$$

where $h_2$ is the height at which the OWT blade is installed; $h_1$ is the height at which the anemometer is installed; $v_{m,h}^{h_1}$ is velocity of wind measured by the anemometer for month $m$ at hour $h$; $z_0$ is the roughness length of the sea which is 0.0002m for open sea. Figure 4 illustrates the average power output of an 8000kW OWT for each hour of every month. It is evident that Alaska has strong wind resources, making it well-suited for OWTs. Texas also generates a significant amount of power from OWTs. However, during July and August (subplots 4.g and 4.h), the power output drops due to lower wind speeds during these months.

Since OWTs are a more advanced technology, it is fair to evaluate multiple candidate sizes of OWT resources to determine which resource is best fit for each region. We have considered, commercially available OWTs at 8000kW, 11000kW, and 14000kW, as well as prototype-stage OWTs at 16000kW and 18000kW, and a concept-stage 22000kW OWT. Figure 5 compares the performance of these different OWT sizes. The total capacity factor over five years is averaged and plotted on the Y-axis, with different OWT sizes on the X-axis. The plot show that the 16000kW OWT performs consistently well across all six regions. Additionally, the 22000kW OWT performs just as well in California, Hawaii and New Jersey and slightly outperforms the 16000kW OWT in Florida. Overall, OWTs in the prototype and concept stages demonstrate strong performance in most regions.

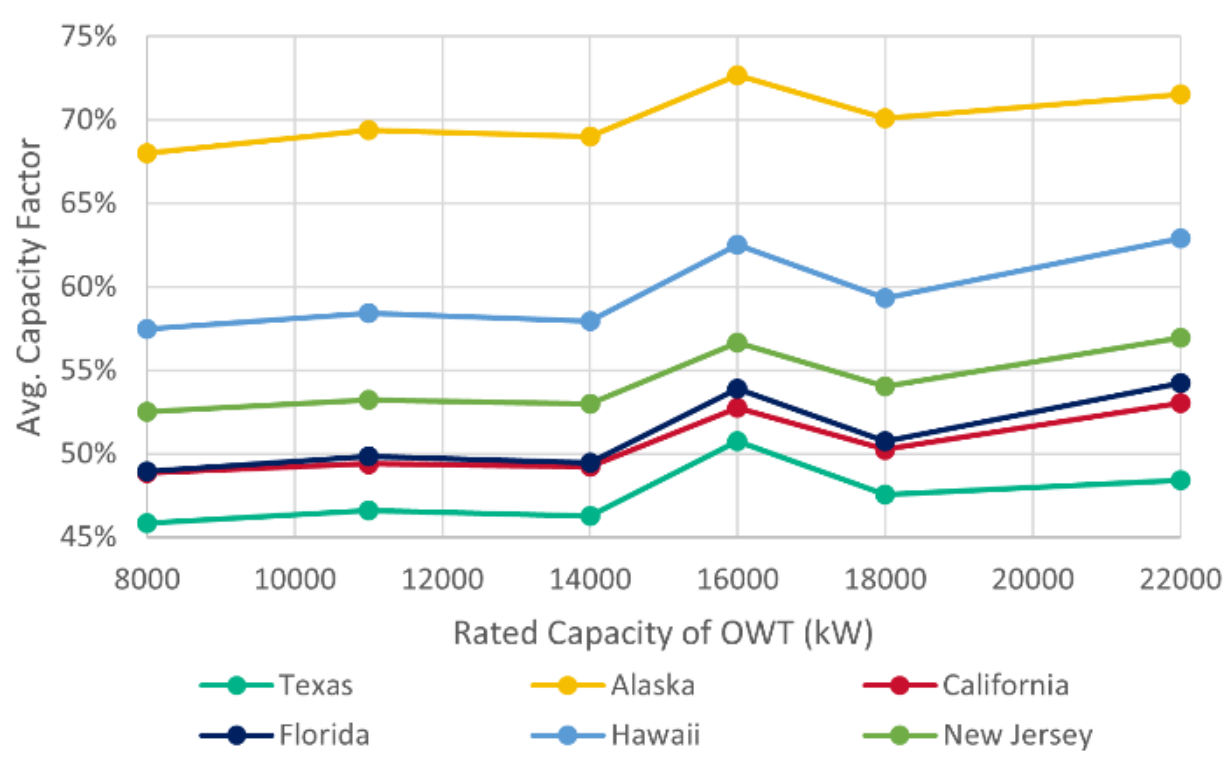

Fig. 5. Average capacity factors of OWTs with different power capacities.

## *2.4. Electrical Power Generation from FPVs*

FPVs are a nascent technology, but there has been an increase in research on the application of this technology for marine applications over the past couple of years. FPVs benefit from the cooling effect of water, which helps reduce overheating of the panels and improves its efficiency. The power generated by the FPVs are calculated using the PVWatts calculator [70].

Figure 6 shows the average power output of a 280kW FPV for each hour of each month. We can see that the regions closer to the earth's equator like Texas, Florida, California and Hawaii receive abundant sunlight year-round making FPVs a strong contender for renewable energy generation in these areas. However, on the other hand, for regions closer to the poles, like Alaska and New Jersey, FPVs perform well mainly in the summer months of June and July (subplots 6.f and 6.g). In contrast, their performance drops significantly in the winter, with the capacity factor reaching only about 18% in December (subplot 6.l).

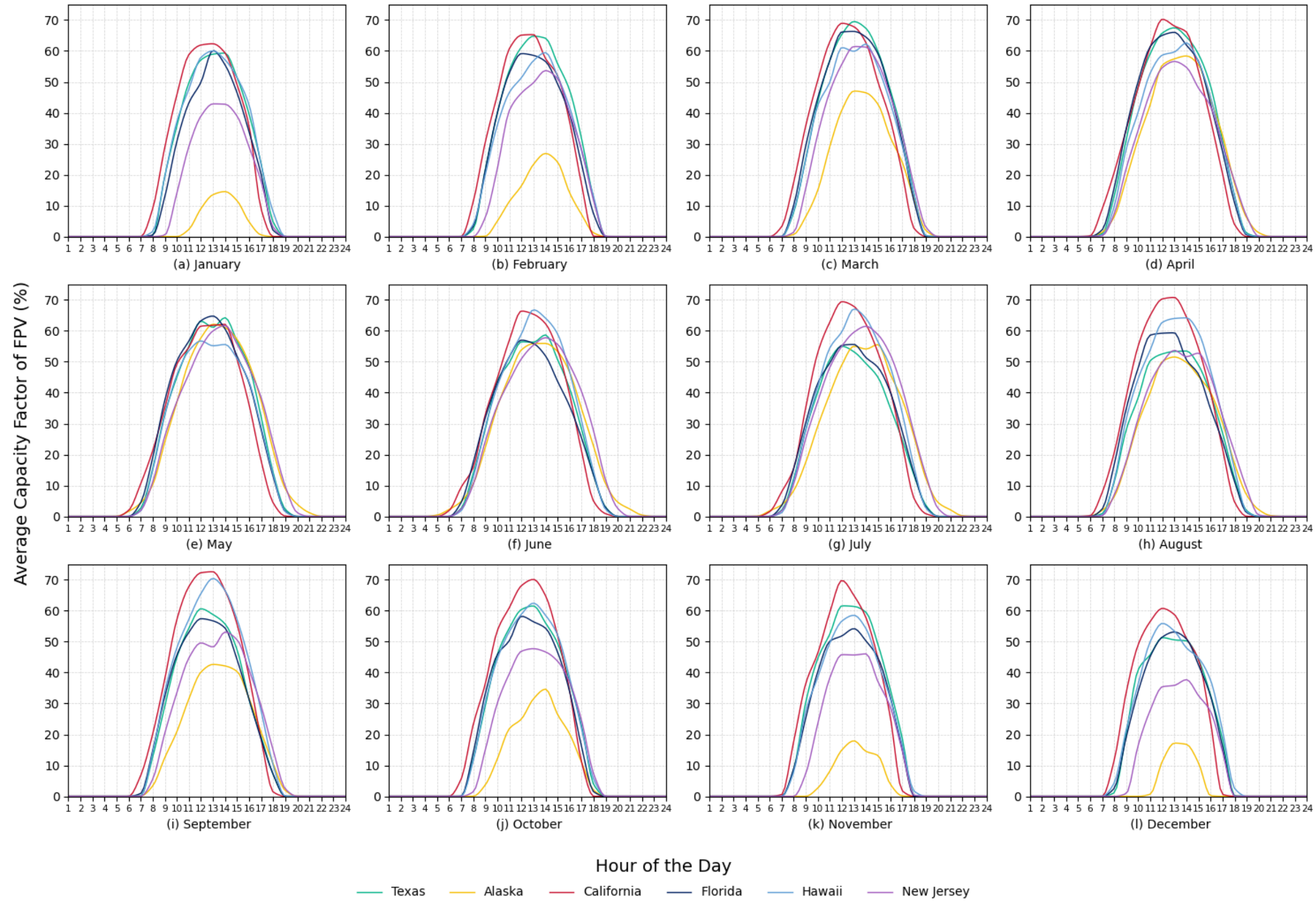

Fig. 6. Average output capacity factor of a single FPV for each hour of each month.

Since the FPVs show promising power generation potential in most regions, it is reasonable to evaluate multiple candidate sizes to determine the best fit for each region. We have considered the commercially available 650kW and 280kW models. Since the power output from FPVs depends on irradiation and temperature, both resources have the same capacity factor. The choice between the two devices would ultimately depend on power consumption and generation needs.

## 3. Subsystem Level Modelling of the Offshore Renewable Energy Microgrid

The proposed REMO sizing model for OREM microgrid is designed to be efficient, flexible, and cost-effective. By using renewable resources and battery energy storage solution. The proposed REMO model for sizing the renewable resources for the OREM microgrid is governed by (4)-(29).

### *3.1. Power Balance Equation*

The power balance equation is a fundamental equation which governs any power system. It ensures that the total power generated within the OREM is always sufficient to meet the total demand at any given time interval. This relationship is described by (4).

$$\sum_{w\in W}\left(N_w \times P_{m,h}^{w}\right) + \sum_{t\in T}\left(N_t \times P_{m,h}^{t}\right) + \sum_{o\in O}\left(N_o \times P_{m,h}^{o}\right) + \sum_{f\in F}\left(N_f \times P_{m,h}^{f}\right) + \left(P_{m,h}^{Disc} - P_{m,h}^{Char}\right) - P_{m,h}^{Curt} = P_{m,h}^{Load} \tag{4}$$

### 3.2. Objective Function

The primary objective of the REMO model is to minimize the overall lifetime cost associated with the proposed OREM system. This is defined by (5).

$$min\ Cost = f\left(C^{Precom}\right) + f\left(C^{Capital}\right) + \left\{f\left(C^{O\&M}\right) \times T_e\right\} + f\left(C^{Decom}\right) \tag{5}$$

The lifetime cost of the system consists of the sum total of the pre-commissioning costs, capital costs, operations and maintenance costs, and decommissioning costs of the entire microgrid system.

The pre-commissioning cost associated with the OREM is the cost incurred during engineering tasks, such as study of the area, design of the OREM, environmental impact studies and consenting process. The pre-commissioning of the OREM is the sum total of the pre-commissioning cost associated with each subsystem of the microgrid. It is defined by (6)-(11).

$$f(C^{Precom}) = f(C_{WEC}^{Precom}) + f(C_{TEC}^{Precom}) + f(C_{OWT}^{Precom}) + f(C_{FPV}^{Precom}) + f(C_{BESS}^{Precom}) \tag{6}$$

$$f(C_{WEC}^{Precom}) = \sum_{w \in W} [N_w \times f(C_w^{Precom})] \tag{7}$$

$$f(C_{TEC}^{Precom}) = \sum_{t \in T} [N_t \times f(C_t^{Precom})] \tag{8}$$

$$f(C_{OWT}^{Precom}) = \sum_{o \in O} [N_o \times f(C_o^{Precom})] \tag{9}$$

$$f(C_{FPV}^{Precom}) = \sum_{f \in F} \left[N_f \times f\left(C_f^{Precom}\right)\right] \tag{10}$$

$$f(C_{BESS}^{Precom}) = E_{BESS}^{total} \times C_{BESS}^{Precom} \tag{11}$$

The capital cost associated with the OREM encompasses the purchase of all the equipment and elements of the microgrid and their installation. It is defined by (12)-(17).

$$f(C^{Capital}) = f\left(C_{WEC}^{Capital}\right) + f\left(C_{TEC}^{Capital}\right) + f\left(C_{OWT}^{Capital}\right) + f\left(C_{FPV}^{Capital}\right) + f\left(C_{BESS}^{Capital}\right) \tag{12}$$

$$f\left(C_{WEC}^{Capital}\right) = \sum_{w \in W} \left[N_w \times f\left(C_w^{Capital}\right)\right] \tag{13}$$

$$f\left(C_{TEC}^{Capital}\right) = \sum_{t \in T} \left[N_t \times f\left(C_t^{Capital}\right)\right] \tag{14}$$

$$f\left(C_{OWT}^{Capital}\right) = \sum_{o \in O} \left[N_o \times f\left(C_o^{Capital}\right)\right] \tag{15}$$

$$f\left(C_{FPV}^{Capital}\right) = \sum_{f \in F} \left[N_f \times f\left(C_f^{Capital}\right)\right] \tag{16}$$

$$f\left(C_{BESS}^{Capital}\right) = E_{BESS}^{total} \times C_{BESS}^{Capital} \tag{17}$$

Operations and maintenance cost associated with the OREM encompasses the costs associated with scheduled maintenance, unscheduled maintenance and insurance costs. It is defined by (18)-(23).

$$f(C^{O\&M}) = f\left(C_{WEC}^{O\&M}\right) + f\left(C_{TEC}^{O\&M}\right) + f\left(C_{OWT}^{O\&M}\right) + f\left(C_{FPV}^{O\&M}\right) + f\left(C_{BESS}^{O\&M}\right) \tag{18}$$

$$f\left(C_{WEC}^{O\&M}\right) = \sum_{w \in W} [N_w \times f(C_w^{O\&M})] \tag{19}$$

$$f\left(C_{TEC}^{O\&M}\right) = \sum_{t \in T} \left[N_t \times f\left(C_t^{O\&M}\right)\right] \tag{20}$$

$$f\left(C_{OWT}^{O\&M}\right) = \sum_{o \in O} [N_o \times f(C_o^{O\&M})] \tag{21}$$

$$f\left(C_{FPV}^{O\&M}\right) = \sum_{f \in F} \left[N_f \times f\left(C_f^{O\&M}\right)\right] \tag{22}$$

$$f\left(C_{BESS}^{O\&M}\right) = E_{BESS}^{total} \times C_{BESS}^{O\&M} \tag{23}$$

Decommissioning costs associated with the OREM encompass expenditure related to dismantling equipment, restoring the site of installation, and complying with relevant environmental regulations. The decommissioning cost associated with the OREM is defined by (24)-(29).

$$f(C^{Decom}) = f(C_{WEC}^{Decom}) + f(C_{TEC}^{Decom}) + f(C_{OWT}^{Decom}) + f(C_{FPV}^{Decom}) + f(C_{BESS}^{Decom}) \tag{24}$$

$$f(C_{WEC}^{Decom}) = \sum_{w \in W} [N_w \times f(C_w^{Decom})] \tag{25}$$

$$f(C_{TEC}^{Decom}) = \sum_{t \in T} [N_t \times f(C_t^{Decom})] \tag{26}$$

$$f(C_{OWT}^{Decom}) = \sum_{o \in O} [N_o \times f(C_o^{Decom})] \tag{27}$$

$$f(C_{FPV}^{Decom}) = \sum_{f \in F} \left[N_f \times f\left(C_f^{Decom}\right)\right] \tag{28}$$

$$f(C_{BESS}^{Decom}) = E_{BESS}^{total} \times C_{BESS}^{Decom} \tag{29}$$

*3.3. Formulation of BESS Model*

Equations (30)-(35) model the BESS in the OREM system. For any given time interval, the energy stored in the BESS is modeled by (30). To reduce the overall stress on the BESS, the battery cells are limited from draining itself completely and from charging itself to 100%. This is modeled by (31). Constraint (32) ensures that the BESS does not charge and discharge at the same instant, while, (33) and (34) regulate the charging and discharging capacity of the BESS. Most of the commercially available BESS capacities are in multiples of 3900 kWh. This is modelled by (35).

$$E_{BESS}^{m,h} - E_{BESS}^{m,h-1} = \left(P_{m,h}^{Char} \cdot \eta_{BESS}^{Char}\right) - \frac{P_{m,h}^{Disc}}{\eta_{BESS}^{Disc}} \quad \forall\, m,h \tag{30}$$

$$SOC_{min} E_{BESS} \leq E_{BESS}^{m,h} \leq SOC_{max} E_{BESS}, \forall\, m,h \tag{31}$$

$$U_{m,h}^{Char} + U_{m,h}^{Disc} \leq 1 , \forall\, m,h \tag{32}$$

$$0 \leq P_{m,h}^{Disc} \leq U_{m.h}^{Disc} \times P_{Max}^{Disc} , \forall\, m,h \tag{33}$$

$$0 \leq P_{m,h}^{Char} \leq U_{m,h}^{Char} \times P_{Max}^{Char} , \forall\, m,h \tag{34}$$

$$E_{BESS}^{total} = E_{BESS} \times N_{BESS} \tag{35}$$

## 4. Deep Neural Network Based Battery Degradation Modelling

Battery degradation refers to the gradual decline of the ability of the BESS to store and deliver energy. This process is inevitable and can result in reduced energy capacity, range, power and overall efficiency of the BESS. Traditionally, BESS degradation costs are estimated to be around 5-10% of the initial capital investment potentially leading to economic inefficiencies due to premature BESS replacement. To address this, battery aging is simulated using MATLAB Simulink and a deep neural network-based battery degradation (DNN-BD) model [52] is trained to calculate the use-based cost associated with BESS degradation.

The training data originates from 945 aging simulations, each representing different combinations of five key inputs: ambient temperature (T), charge/discharge rate (C Rate), state of charge (SOC), depth of discharge (DOD) and state of health (SOH). Each input vector corresponds to a single output value which represents the percentage of battery degradation with respect to the SOH of that cycle. The simulations continue until the BESS's capacity drops to 80% of its rated value, a commonly accepted threshold marking the end of life of the BESS. These simulations produce cycle-by-cycle data from which the BESS degradation is calculated. To enhance training performance, the degradation data undergoes three preprocessing methods; raw, smoothed and linearly regressed. Smoothed data is found to improve the DNN training accuracy without affecting long term degradation trends. Data normalization is applied before splitting the dataset into 80% training and 20% validation subsets.

The structure of the deep neural network consists of a fully connected feedforward model with 5 input neurons, a first hidden layer with 20 neurons, a second hidden layer with 10 neurons, and an output layer with 1 neuron. The hidden layers use the ReLU activation function, followed by linear activation in the output layer to predict degradation per cycle. The output of the neural network provides the estimated BESS degradation that is then used to calculate the associated degradation cost. The network is trained using mini batch gradient descent, and a dynamic learning rate that decreases over training epochs.

The proposed DNN-BD was tested on a typical grid-connected microgrid with renewable energy sources, including wind turbines, roof top photovoltaic panels and lithium-ion BESS. Real world data was obtained from Pecan Street dataport, while electricity price data was sourced from ERCOT. The model was used to quantify short term BESS degradation over a 24 hour horizon within a realistic operational scenario. This DNN-BD model achieved high prediction accuracy, with cumulative degradation error difference below 1.5% compared to raw data. This reliable performance supports its potential integration into the long term REMO model.

The objective function of REMO needs to be updated to include the equivalent usage-based BESS degradation. The updated objective function is represented by (36):

$$min\ Cost = f\left(C^{Precom}\right) + f\left(C^{Capital}\right) + \left\{f\left(C^{O\&M}\right) \times T_e\right\} + f\left(C^{Decom}\right) + f\left(C^{BESS\ Deg}\right) \quad (36)$$

where $C^{BESS\ Deg}$ is the cost associated with BESS degradation that is calculated by the DNN-BD model.

This DNN-BD model is highly nonlinear and non-convex making the optimization model computationally demanding. To address this challenge, a decoupled heuristic algorithm is used to solve this problem in an iterative process as shown in Figure 7.

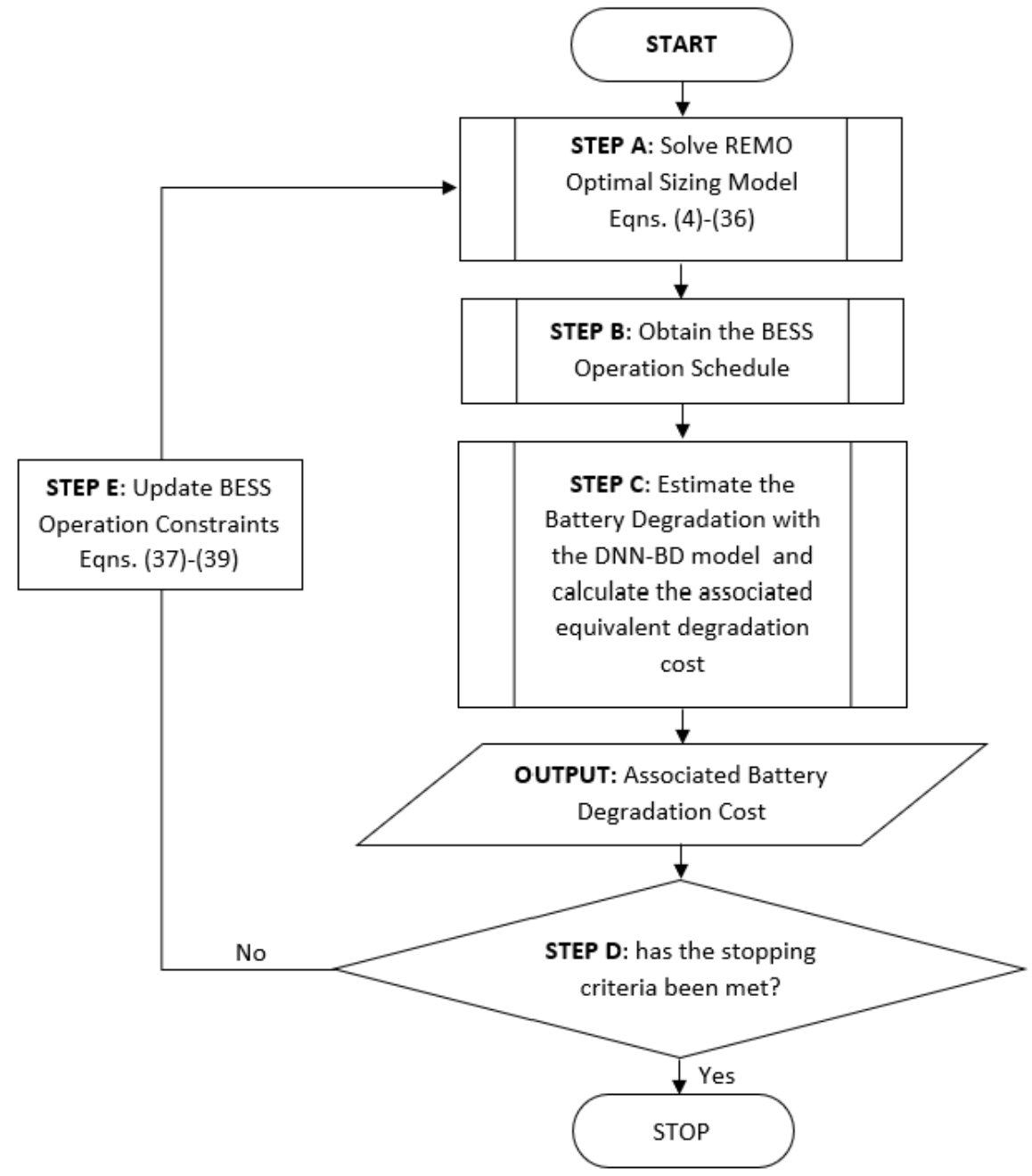

Fig. 7. Flowchart of the decoupled heuristic algorithm.

In Step A, the REMO optimization model is solved. In the first iteration, no additional constraints are applied to the operation of the BESS. In the subsequent iterations, the usage of the BESS is limited by applying the constraints introduced in Step E. In step B, the BESS operation schedule from the optimization model is analyzed, and operation time intervals are grouped and averaged into distinct cycles. If the BESS remains in the same state (charging or discharging) for consecutive time intervals, those intervals are aggregated into a single charging or discharging cycle. In step C, the cost associated with BESS degradation is calculated using the DNN-BD model. In step D, the stopping criteria is evaluated. If met, the algorithm terminates; otherwise, it proceeds to Step E. In step E, additional constraints on battery operation are introduced to limit BESS usage, reducing degradation and, in turn, minimizing the BESS degradation costs. Constraints (37) and (38) are designed to limit the maximum power that is charged/discharged by the BESS, whereas (39) reduces the sum of the overall battery output by forcing it to be lesser than the previous iteration.

$$P_{m,h.i}^{Char} \leq P_{Max,i}^{Char}\left(1-\alpha\right)^{i-1} \tag{37}$$

$$P_{m,h.i}^{Disc} \leq P_{Max,i}^{Disc}\left(1-\alpha\right)^{i-1} \tag{38}$$

$$\sum_{\substack{m \in M \\ h \in H}}\left(P_{m,h.i}^{Char}+P_{m,h.i}^{Disc}\right) \leq \left(1-\alpha\right) \times \sum_{\substack{m \in M \\ h \in H}}\left(P_{m,h.(i-1)}^{Char}+P_{m,h.(i-1)}^{Disc}\right) \tag{39}$$

where $\alpha$ is a preset control parameter referred to as the BESS usage control factor.

## 5. Updated Subsystem Level Modelling of the Offshore Renewable Energy Microgrid

Figures 2 and 3 indicate that WECs and TECs do not perform well in certain regions. However, relying solely on visual observation may not be sufficient to justify completely excluding these resources from the optimization model. To make a better-informed decision, the levelized cost of energy (LCOE) for each resource in these regions is analyzed. The LCOE is calculated by dividing the total lifetime costs associated with a generation resource by the total energy it generates over its lifetime. Table 2 presents the LCOE for each generation resource across different regions.

Table 2. LCOE of the four offshore renewable resources considered for the offshore microgrid

| Asset | ***Levelized Cost of Energy (¢/kWh)*** | | | | | |
|---|---|---|---|---|---|---|
| | ***TX*** | ***AK*** | ***CA*** | ***FL*** | ***HI*** | ***NJ*** |
| ***WEC*** | 172.5 | 44.2 | 87.6 | 109.5 | 67.1 | 155.1 |
| ***TEC*** | 1653.3 | 448.7 | - | 83.2 | 3427.2 | 33527 |
| ***OWT*** | 6.3 | 4.5 | 6.2 | 6.0 | 5.2 | 5.7 |
| ***FPV*** | 7.8 | 12.4 | 7.3 | 8.1 | 7.7 | 9.0 |

The table clearly shows that OWTs and FPVs have significantly lower LCOEs compared to TECs and WECs. Given that WECs and TECs exhibit consistently high LCOEs across all six regions, it is reasonable to exclude them from the optimization model. Removing these resources would help reduce computational complexity and processing time. Henceforth, the base model will be referred to as the REMO-classic model and the updated model will be referred to as REMO-simplified model.

The updated subsystem equations are given by (40)-(43):

$$f(C^{Precom}) = f(C_{OWT}^{Precom}) + f(C_{FPV}^{Precom}) + f(C_{BESS}^{Precom}) \tag{40}$$

$$f(C^{Capital}) = f\left(C_{OWT}^{Capital}\right) + f\left(C_{FPV}^{Capital}\right) + f\left(C_{BESS}^{Capital}\right) \tag{41}$$

$$f(C^{O\&M}) = f\left(C_{OWT}^{O\&M}\right) + f\left(C_{FPV}^{O\&M}\right) + f\left(C_{BESS}^{O\&M}\right) \tag{42}$$

$$f(C^{Decom}) = +f(C_{OWT}^{Decom}) + f(C_{FPV}^{Decom}) + f(C_{BESS}^{Decom}) \tag{43}$$

## 6. Case Studies

The case studies are organized into 6 subsections. The first subsection examines the optimal sizing of the OREM for oil and gas platforms across the six regions using the REMO-classic model. These platforms are chosen as a case study because their power consumption remains relatively constant throughout the year. The second case study analyzes how the OREM sizing would change when the wind speeds are varied by ±10%. The third case study focuses on the same platforms, but this time using the REMO-simplified model to evaluate the impact of reduced constraints. The fourth case study examines the OREM sizing for a coastal community, specifically a large coastal town, where power consumption varies seasonally. The fifth and sixth case studies evaluate the impact of cost variations on the competitiveness of WECs and TECs, identifying the cost threshold at which they could be considered competitive renewable energy resources. The proposed REMO models are implemented on Python using Pyomo [71], and Gurobi [72] is employed as the optimizer solver. The model considers W = {750}; T = {300}; F = {280,650}; and O = {8000,11000,14000,16000,18000,22000}.

*6.1. OREM for Offshore Oil and Gas Platform using REMO-classic model*

There are different types of offshore oil and gas platforms. The electrical power consumption of these platforms can vary substantially from 6MW [73] to 250MW [74] depending on their size and distance from the shore. Fig. 8 shows the electrical load profile of a test bed platform. Unlike residential loads, the electrical power consumption does not fluctuate much with the seasons. This is primarily because most of the power is used for mining and oil processing activities which constitutes 80% of the load which remains relatively the same throughout the year [75], [76].

Fig. 8. Load profile for the offshore platform

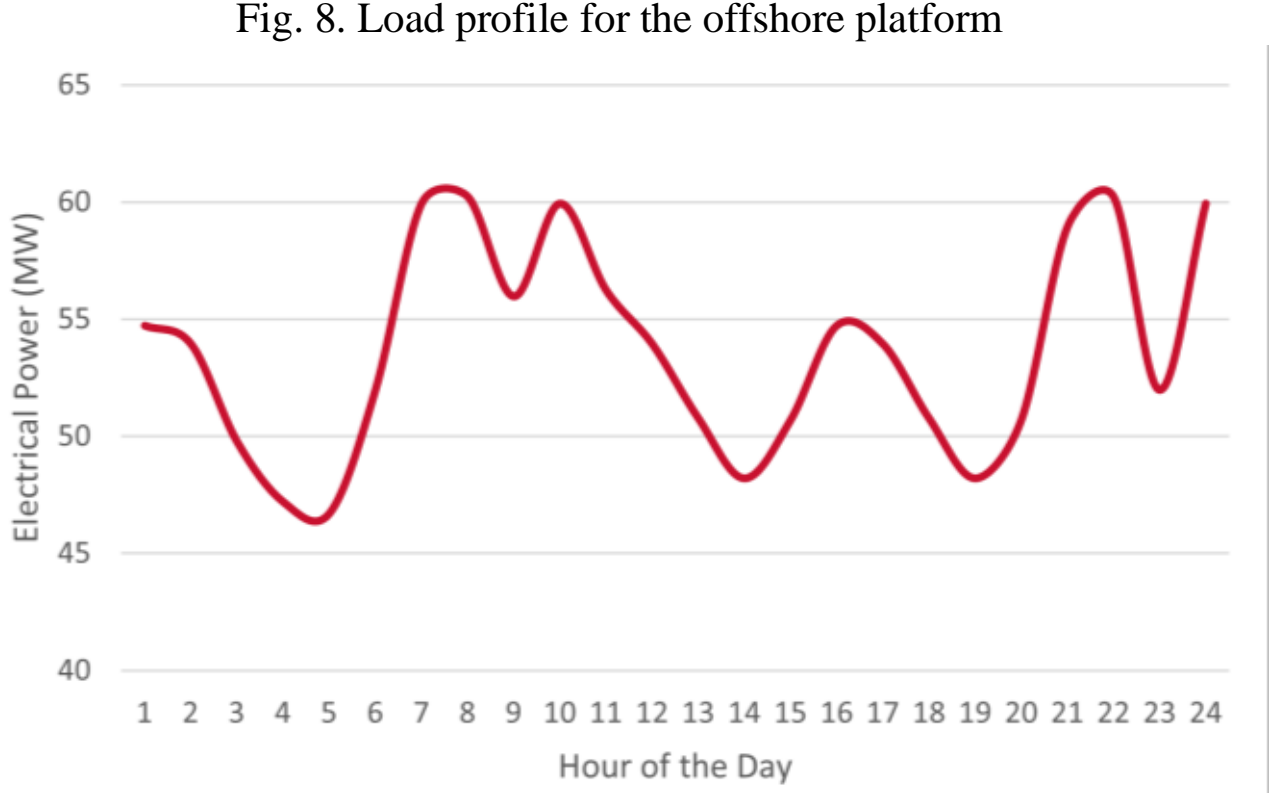

Table 3 presents the optimized lifetime costs of the proposed OREM system across 6 different regions using the REMO-classic model. This cost was calculated for the total electricity consumption of 1290MWh per day by an offshore platform over a 20-year period. The BESS usage control factor is set at 0.001. The BESS operation is constrained by (37)-(39).

Table 3. Optimized lifetime cost and quantities of energy generation resources for offshore oil and gas platforms

| ***Asset\Region*** | ***TX*** | ***AK*** | ***CA*** | ***FL*** | ***HI*** | ***NJ*** |
|---|---|---|---|---|---|---|
| ***Cost (M$)*** | 741.7 | 478.2 | 647.2 | 728.9 | 592.1 | 724.3 |
| ***WEC*** | 0 | 0 | 0 | 0 | 0 | 0 |
| ***TEC*** | 0 | 0 | 0 | 0 | 0 | 0 |
| ***8MW OWT*** | 1 | 1 | 0 | 0 | 0 | 0 |
| ***11MW OWT*** | 0 | 0 | 0 | 0 | 0 | 0 |
| ***14MW OWT*** | 0 | 0 | 0 | 0 | 0 | 0 |
| ***16MW OWT*** | 6 | 4 | 2 | 0 | 2 | 1 |
| ***18MW OWT*** | 0 | 0 | 0 | 0 | 1 | 0 |
| ***22MW OWT*** | 0 | 0 | 3 | 5 | 2 | 4 |
| ***650kW FPV*** | 9 | 1 | 1 | 0 | 0 | 0 |
| ***280kW FPV*** | 1 | 0 | 0 | 1 | 0 | 22 |
| ***BESS (kWh)*** | 171,600 | 74,100 | 101,400 | 97,500 | 66,300 | 117,000 |

The table shows that, despite considering the same oil and gas platform, the optimal lifetime cost varies by region. Alaska has the lowest cost due to its strong and consistent winds, allowing for higher power generation from the OWT, reducing the need for additional resources. In contrast, New Jersey has the highest lifetime cost because of weaker winds during certain months, making the OREM system more reliant on FPVs and BESS.

Another interesting fact to note is that, if this particular offshore oil and gas platform is operated solely on diesel, it would generate approximately 332,289 tonnes of $CO_2e$ over its lifetime. In 2017, the High Level Commission on Carbon Prices estimated a direct carbon prices of \$50-100 per tonne of $CO_2e$ by 2030 [77]. If these carbon credit rates are applied, the OREM becomes economically competitive across all studies regions.

Table 4 shows a comparison for different values of the BESS usage control factor for Texas when the BESS is constrained by (37)-(39).

Table 4. Optimized lifetime cost and quantities of energy generation resources for different usage control parameters

| *Assets* | *No constraint* | *α=0.001* | *α=0.01* |
|---|---|---|---|
| ***Lifetime Cost (M\$)*** | **760.6** | **741.7** | **743.3** |
| *BESS Degradation Cost (M\$)* | 53.5 | 30.6 | 30.7 |
| *Lifetime Cost excluding BESS Degradation (M\$)* | 707.1 | 711.1 | 712.6 |
| ***8MW OWT*** | 1 | 1 | 0 |
| ***11MW OWT*** | 0 | 0 | 1 |
| ***14MW OWT*** | 0 | 0 | 0 |
| ***16MW OWT*** | 6 | 6 | 6 |
| ***18MW OWT*** | 0 | 0 | 0 |
| ***22MW OWT*** | 0 | 0 | 0 |
| ***650kW FPV*** | 7 | 9 | 0 |
| ***280kW FPV*** | 0 | 1 | 3 |
| ***BESS (kWh)*** | 171,600 | 171,600 | 167,700 |

The table shows that without any constraint on BESS operation, the cost due to battery degradation is \$53.5 million. Applying a small constraint limiting the BESS operating range can reduce degradation costs by 42%, despite an increase in the assets connected to the OREM. When a larger constraint is applied, the size of the BESS decreases, resulting in a \$17.1 million reduction in the overall lifetime cost.

### *6.2. OREM for Offshore Oil and Gas Platform for varied wind speed*

In the previous case study, we demonstrated that powering the same type of offshore oil and gas platform can require significantly different OREM configurations depending on the local meteorological conditions. These differences arise because the sizing of the renewable generation resources and energy storage system must be tailored to match the long term energy yield potential of the site. To further explore the sensitivity of the system to resource variability, Table 5 presents an additional case study for the Texas site, where wind speed is assumed to vary by ±10% from the baseline values.

When wind speed is increased by 10%, the corresponding power output from the OWTs rise significantly due to the cubic relationship between wind speed and OWT power production. This enhanced generation capacity reduces the OWTs required in the OREM and lower the required BESS capacity, ultimately decreasing the lifetime investment cost by 16%. Conversely, when wind speed is reduced by 10%, the OWTs power output drops sharply by 27%, requiring additional renewable generation and BESS capacity . This results in a notable increase in expenditure, with lifetime investment costs rising by 33%.

Table 5. Optimized lifetime cost and quantities of energy generation resources for 10% wind speed variation

| *Assets* | *Wind speed increased by 10%* | *Baseline wind speeds* | *Wind speed decreased by 10%* |
|---|---|---|---|
| ***Lifetime Cost (M\$)*** | **618.5** | **741.7** | **986.9** |
| *BESS Degradation Cost (M\$)* | 58.4 | 30.6 | 59.1 |
| *Lifetime Cost excluding BESS Degradation (M\$)* | 560.1 | 711.1 | 927.7 |
| ***8MW OWT*** | 0 | 1 | 0 |
| ***11MW OWT*** | 0 | 0 | 0 |
| ***14MW OWT*** | 0 | 0 | 0 |
| ***16MW OWT*** | 5 | 6 | 9 |
| ***18MW OWT*** | 0 | 0 | 0 |
| ***22MW OWT*** | 0 | 0 | 0 |
| ***650kW FPV*** | 0 | 9 | 0 |
| ***280kW FPV*** | 5 | 1 | 9 |
| ***BESS (kWh)*** | 167,700 | 171,600 | 167,700 |

This scenario analysis provides insight into how even modest changes in wind resource potential can have a substantial impact on optimal sizing and overall project economics.

### *6.3. OREM for Offshore Oil and Gas Platforms using REMO-simplified model*

In the previous model, the input dataset considered 12 typical days -representing an average day for each month- to represent the power generation for an entire year. In this case study, the resolution of the dataset was increased to consider for 365 unique days to represent an entire year. These datasets were analyzed for 3 regions -Texas, Alaska, and California - by both the REMO-classic and REMO-simplified model. As shown in Figure 9, the REMO-simplified model consistently requires lesser computation time compared to the REMO-classic model. For Texas and California, the REMO-simplified model reduced the computation time by 16%. In Alaska, the computation time reduced by 25%. This is expected because in the REMO-classic model, the WEC produces sufficient electrical power to be considered a candidate generation resource.

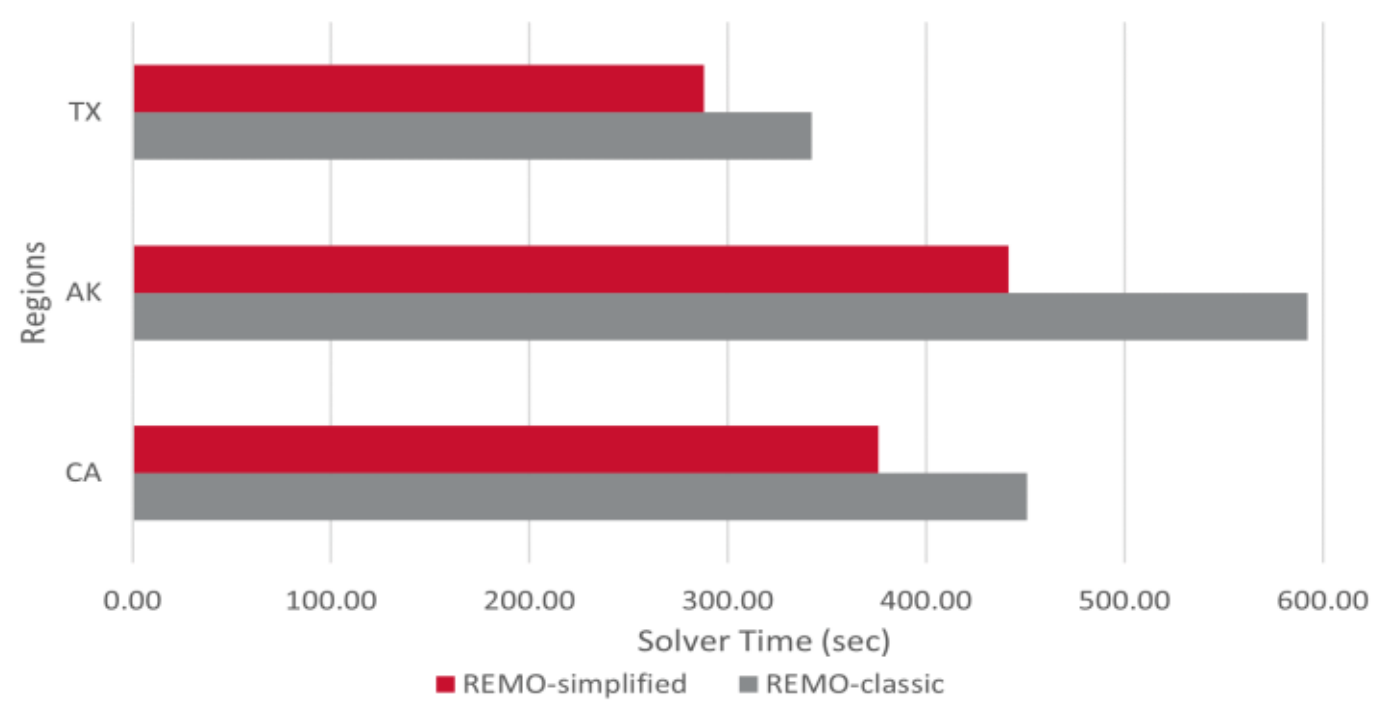

Fig. 9. Computational time comparison between REMO-classic and REMO-simplified model

### *6.4. OREM for Large Coastal Town using REMO simplified model*

In the previous case studies, the load profile was assumed to be constant throughout the year, with the OREM size determined solely by the power generation potential in each region and the cost of each asset. In this case study, we're using the load profile of a large coastal town, shown in Fig. 10, where power consumption is higher from May to October. While this profile may not reflect consumption patterns in all six regions, the focus of this research is on the sizing model, so we assume the same consumption profile applies across all regions.

The optimized lifetime costs of the proposed OREM system across 6 different regions are presented in Table 6. The BESS usage control factor is set at 0.01 in (37)-(39).

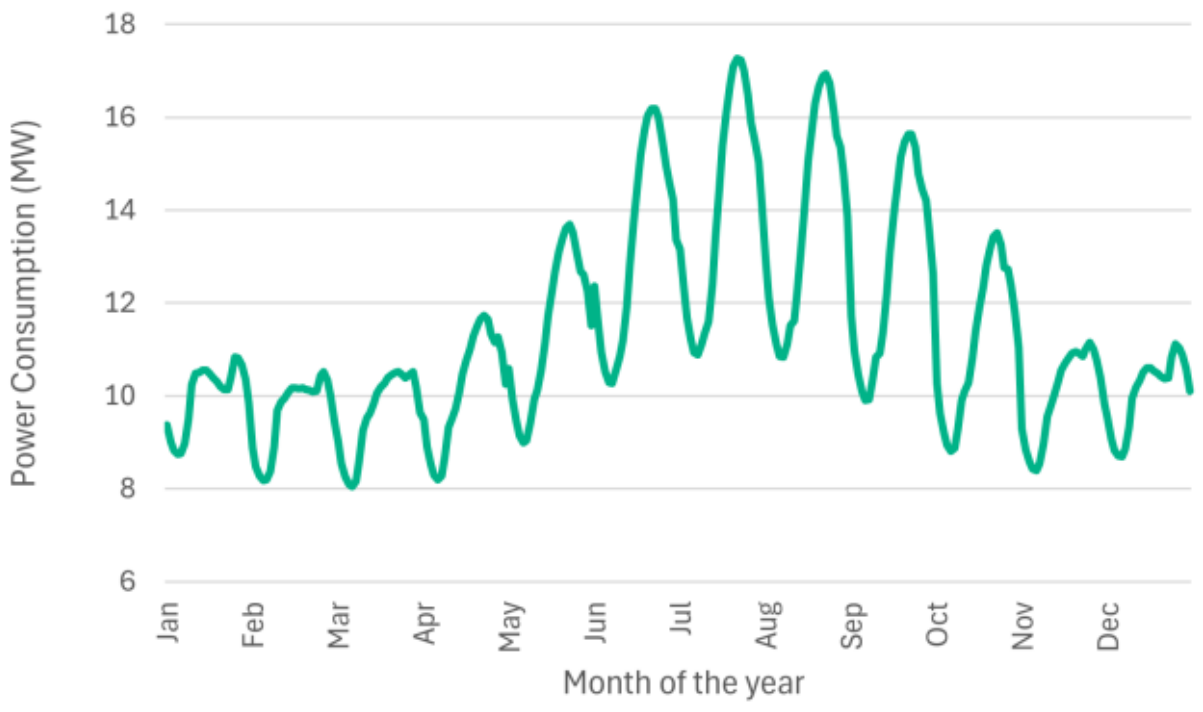

Fig. 10. Load profile for the coastal community

Table 6. Optimized lifetime cost and quantities of energy generation resources for a large coastal town

| ***Asset\Region*** | ***TX*** | ***AK*** | ***CA*** | ***FL*** | ***HI*** | ***NJ*** |
|---|---|---|---|---|---|---|
| ***Cost (M$)*** | 129.7 | 108.6 | 131.7 | 149.3 | 136.5 | 148.3 |
| ***8MW OWT*** | 0 | 0 | 0 | 0 | 0 | 0 |
| ***11MW OWT*** | 1 | 0 | 0 | 0 | 0 | 0 |
| ***14MW OWT*** | 0 | 1 | 0 | 0 | 0 | 1 |
| ***16MW OWT*** | 0 | 0 | 1 | 1 | 1 | 0 |
| ***18MW OWT*** | 0 | 0 | 0 | 0 | 0 | 0 |
| ***22MW OWT*** | 0 | 0 | 0 | 0 | 0 | 0 |
| ***650kW FPV*** | 0 | 0 | 0 | 0 | 1 | 0 |
| ***280kW FPV*** | 3 | 1 | 1 | 0 | 10 | 16 |
| ***BESS (kWh)*** | 66,300 | 31,200 | 42,900 | 50,700 | 39,000 | 46,800 |

From May to October, when power consumption is higher in the coastal area, Texas, Florida, and New Jersey experience weaker winds, leading to lower power generation from wind resources. As a result, the BESS size is larger, causing more frequent charging and discharging, which accelerates battery degradation and increases lifetime costs. Hawaii, however, presents a different trend. The REMO model relies heavily on FPVs and BESS due to a significant drop in the OWT's output from May to August, with the average capacity factor declining to 52%. This period also coincides with increased energy demand. Meanwhile, FPVs reach their peak performance, achieving a 68% capacity factor, making them the preferred renewable resource. However, since FPV generation is limited to daylight hours, frequent BESS cycling occurs, further accelerating battery degradation and increasing associated costs.

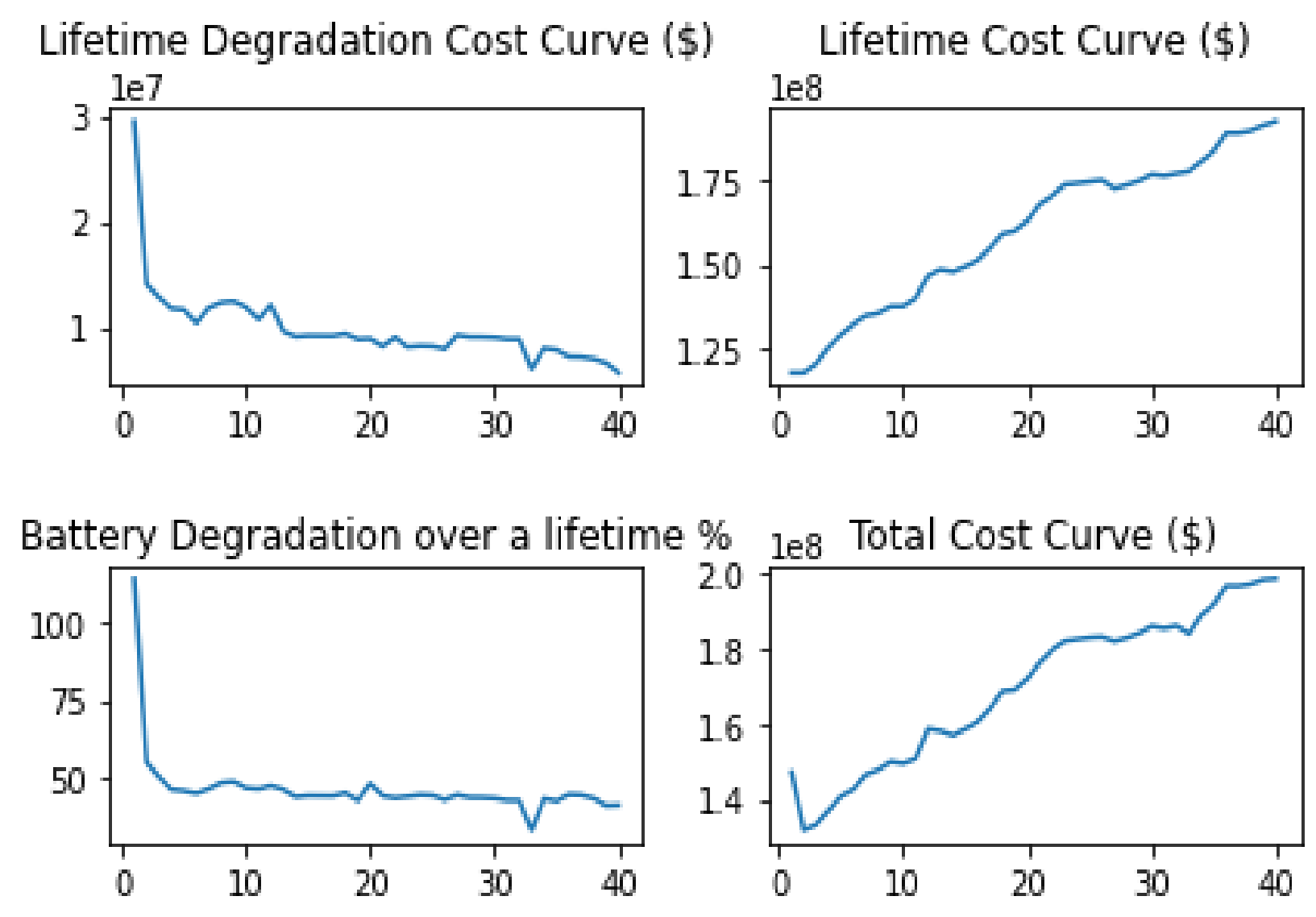

Fig. 11. Results from the iterative REMO-DNN-BD method for California.

Figure 11 displays the results from the DNN-BD model for California. It includes the lifetime degradation cost curve, lifetime cost curve, BESS degradation in percentage over time, and the total cost curve. The figure shows that the total cost curve reaches its lowest point at the third iteration. At this point, the total cost is reduced by 7.7%, demonstrating that considering the DNN-BD model significantly lowers the total cost. As the constraint on BESS operation tightens, the lifetime degradation cost continues to decrease. However, since the lifetime cost curve keeps increasing, the total cost curve rises after hitting its minimum at the third iteration.

### *6.5. Sensitivity Analysis on Cost associated with WEC*

Given the potential of WECs in Alaska, a sensitivity analysis is conducted on their lifetime cost to determine the price point at which they become a competitive renewable resource for integration into the OREM. Table 7 and Figure 12 clearly illustrate this relationship. The analysis indicates that WECs begin to integrate into the OREM once their cost is reduced to 13% of the current level. At 7% of the current cost, WECs supply 95% of the total load for the oil and gas platform. However, during months such as June and July, when WEC power output is significantly lower, the OREM must be supplemented with floating photovoltaic (FPV) systems and large-capacity battery energy storage systems (BESS) to ensure a stable energy supply.

Table 7. Sensitivity Analysis of the Cost associated with WEC in Alaska

| ***Assets*** | ***Percentage of Existing Cost*** | | | | |
|---|---|---|---|---|---|
| | ***100-14%*** | ***13%*** | ***11%*** | ***9%*** | ***7%*** |
| ***Cost (M$)*** | 478.2 | 478.2 | 472.9 | 467.6 | 433.5 |
| ***750kW TEC*** | 0 | 2 | 17 | 69 | 260 |
| ***8MW OWT*** | 1 | 0 | 0 | 0 | 0 |
| ***16MW OWT*** | 4 | 3 | 4 | 3 | 0 |
| ***22MW OWT*** | 0 | 1 | 0 | 0 | 0 |
| ***650kW FPV*** | 0 | 1 | 1 | 0 | 0 |
| ***280kW FPV*** | 1 | 0 | 0 | 4 | 9 |
| ***BESS (kWh)*** | 74,100 | 81,900 | 85,800 | 101,400 | 159,900 |

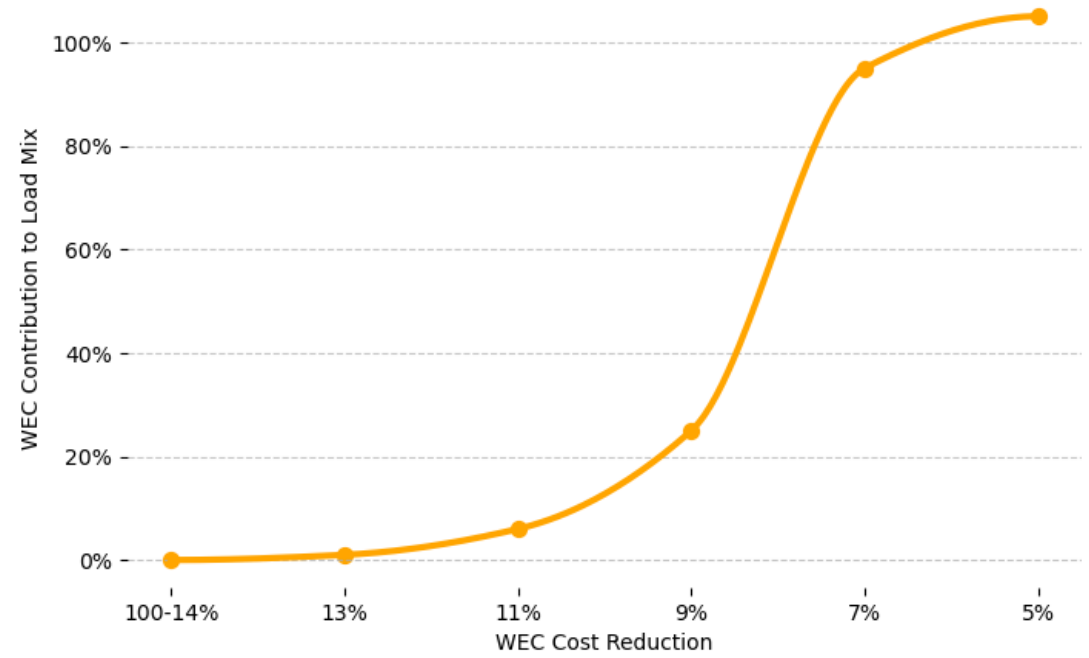

Fig. 12. Sensitivity to WEC Lifetime Cost Reduction

### *6.6. Sensitivity Analysis on Cost associated with TEC*

Given the promising potential of TECs in Florida, a sensitivity analysis is conducted on their lifetime cost to determine the price point at which they become a competitive renewable resource for integration into the OREM. Table 8 and Fig. 13 illustrate the relationship between cost reduction and the extent to which TECs are incorporated into the OREM. TECs remain economically unviable until their overall cost is reduced to 10% of the current level. However, once this threshold is surpassed, they emerge as a competitive energy source. At 7% of the current cost, TECs become a significant contributor to the energy mix for oil and gas platforms, and at 5%, they serve as the primary energy generation resource for the OREM.

Table 8. Sensitivity Analysis of the Cost associated with TEC in Florida

| ***Assets*** | ***Percentage of Existing Cost*** | | | | | | |
|---|---|---|---|---|---|---|---|
| | ***100-11%*** | ***10%*** | ***9%*** | ***8%*** | ***7%*** | ***6%*** | ***5%*** |
| ***Cost (M$)*** | 727.3 | 725.1 | 718.2 | 693.3 | 644.3 | 585.2 | 513.4 |
| ***300kW TEC*** | 0 | 60 | 287 | 472 | 743 | 7068 | 1471 |
| ***16MW OWT*** | 0 | 1 | 1 | 0 | 0 | 0 | 0 |
| ***22MW OWT*** | 5 | 4 | 3 | 3 | 2 | 1 | 0 |

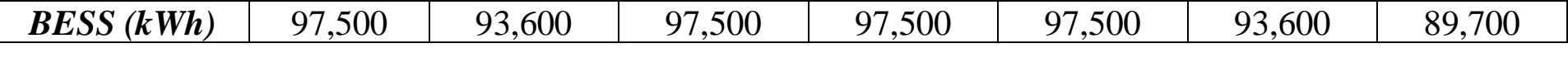

| ***BESS (kWh)*** | 97,500 | 93,600 | 97,500 | 97,500 | 97,500 | 93,600 | 89,700 |
|---|---|---|---|---|---|---|---|

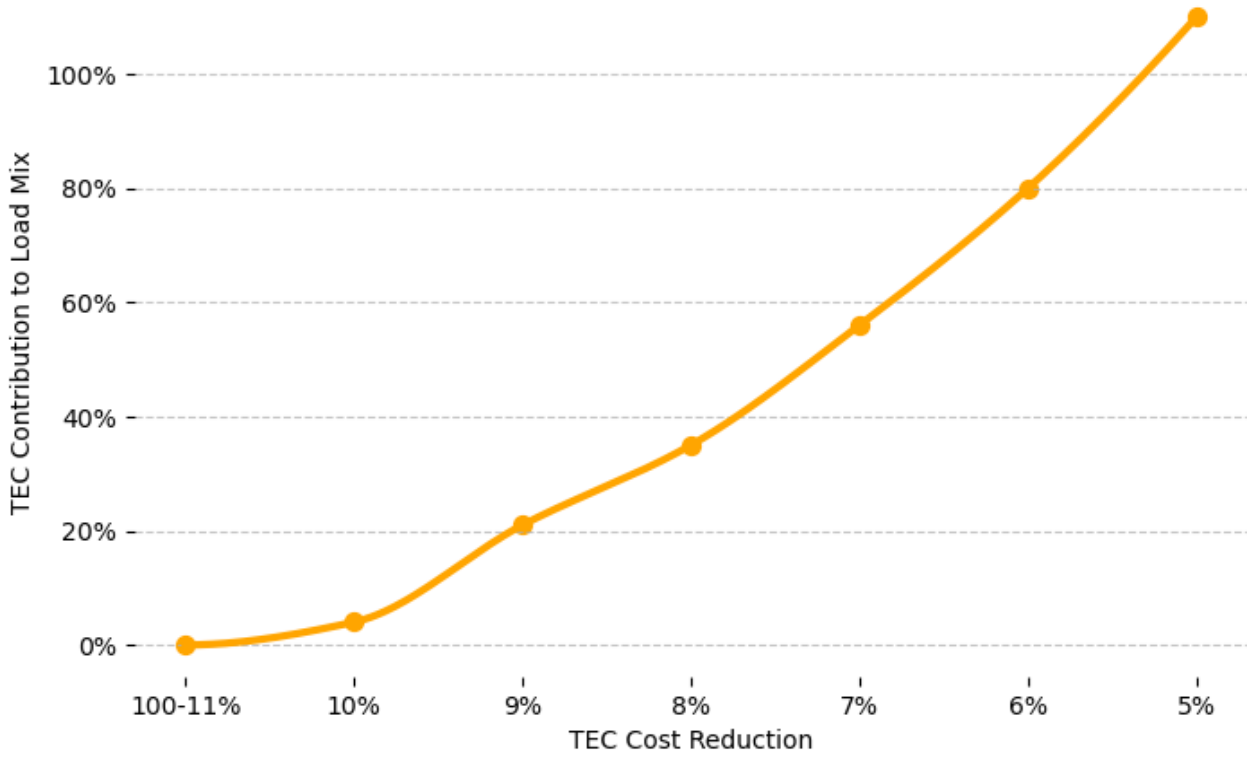

Fig. 13. Sensitivity Analysis of TEC Lifetime Cost Reduction

## 7. Conclusions

The ocean offers immense potential as a renewable energy resource, capable of reducing the strain on traditional mainland grids by powering offshore and coastal communities. However, fully utilizing this potential is hindered by technological, logistical, infrastructural, and regulatory challenges. Continued research and development are essential to design OREMs that are efficient, cost-effective, and reliable.

The proposed REMO sizing model for the OREM microgrid is designed to be efficient, flexible, and cost-effective by utilizing renewable energy sources and smart energy storage solutions. The proposed REMO models for sizing the renewable resources for the OREM is governed by (4)-(36) and is integrated with a DNN-BD model. The DNN-BD model iteratively calculates BESS degradation and, when coupled with the REMO models, helps identify the optimal setup for minimizing the total cost of an OREM.

This paper analyzed four different types of renewable energy resources across six regions. The findings indicate that OWTs and FPVs performed well in most regions due to their technological maturity. In contrast, TECs and WECs, which are still in the early stages of development, did not perform as effectively across all regions. The study also examined the OREM microgrid, which integrated these renewable generation resources alongside BESS to supply electricity to two different types of offshore communities: an offshore oil and gas platform and a coastal community.

Test results show that the proposed REMO-DNN-BD models can reduce the overall degradation cost by about 42% and the total cost by 7.7%. Sensitivity analysis of the costs associated with WECs and TECs reveals that these resources could become competitive once their costs are reduced to 13% and 10% respectively in specific regions.

This paper demonstrates the effectiveness of the proposed REMO models, coupled with the DNN-BD model, in reducing both BESS degradation costs and the overall total cost.

## 8. Future Works

In this study, the REMO microgrid is designed to supply electricity to two different types of offshore communities: an offshore oil and gas platform and a coastal community. However, in practice, many platforms are located close to one another. Designing a larger microgrid to serve a cluster of platforms could be a more efficient solution. Currently, this research only considers BESS as the storage resource. However, we can see that for certain months, both OWTs and FPVs performs poorly making the OREM primarily dependent on the BESS to supply its energy needs. To enhance resilience, the storage solution could be diversified by incorporating two types of storage solutions: BESS as a short-term storage solution which handles daily load fluctuations and a hydrogen energy storage system or similar long-term storage solution to cater to the larger seasonal variations. The REMO model is an extendable model and can be further enhanced to a multi-objective model that integrates carbon credit market and additional resiliency constraints to better capture the OREM's resilience under adverse weather conditions. Furthermore, the incorporation of wake effect modelling could be considered in future works to improve the accuracy of wind turbine performance prediction and optimize the OREM design accordingly.

## Acknowledgments

This project was paid for [in part] with federal funding from the Department of the Treasury through the State of Texas under the Resources and Ecosystems Sustainability, Tourist Opportunities, and Revived Economies of the Gulf Coast States Act of 2012 (RESTORE Act). The statements, findings, conclusions, and recommendations are those of the author(s) and do not necessarily reflect the views of the State of Texas or the Department of the Treasury.